\documentclass[journal]{IEEEtran}

\usepackage{graphicx}
\usepackage{amsmath}
\usepackage[latin1]{inputenc}

\begin{document}

\title{Online Pattern Recognition\\for the ALICE High Level Trigger}

%
%
\author{V.~Lindenstruth$^1$, C.~Loizides$^{2,3}$, D.~Roehrich$^4$, B.~Skaali$^5$, T.~Steinbeck$^1$, R.~Stock$^2$,\\ H.~Tilsner$^1$, K.~Ullaland$^4$, A.~Vestb{\o}$^4$ and T.~Vik$^5$ for the ALICE Collaboration%
\thanks{$^1$Kirchhoff Institut für Physik, Im Neuenheimer Feld 227, D-69120 Heidelberg, Germany}%
\thanks{$^2$Institut für Kernphysik Frankfurt, August-Euler-Str. 6, D-60486 Frankfurt am Main, Germany}%
\thanks{$^3$Corresponding author, email: loizides@ikf.uni-frankfurt.de}%
\thanks{$^4$Department of Physics, University of Bergen, Allegaten 55, N-5007 Bergen, Norway}%
\thanks{$^5$Department of Physics, University of Oslo, P.O.Box 1048 Blindern, N-0316 Oslo, Norway}%
}%

\maketitle

\begin{abstract}

The ALICE High Level Trigger has to process data online, in order to select interesting (sub)events, or to compress data efficiently by modeling techniques. 
Focusing on the main data source, the Time Projection Chamber (TPC), we present two pattern recognition methods under investigation: a sequential approach (\emph{cluster finder} and \emph{track follower}) and an iterative approach (\emph{track candidate finder} and \emph{cluster deconvoluter}). We show, that the former is suited for pp and low multiplicity PbPb collisions, whereas the latter might be applicable for high multiplicity PbPb collisions, if it turns out, that more than 8000 charged particles would have to be reconstructed inside the TPC. Based on the developed tracking schemes we show, that using modeling techniques a compression factor of around 10 might be achievable.




\end{abstract}


%


\section{Introduction}
\label{introduction}

The ALICE Experiment~\cite{alice} at the upcoming Large Hadron Collider at CERN will investigate PbPb collisions at a center of mass energy of about 5.5 TeV per nucleon pair and pp collisions at 14 TeV. Its main tracking detector, the Time Projection Chamber (TPC), is readout by 557568 analog-to-digital channels (ADCs), producing a data size of $\sim$75 MByte per event for central PbPb collisions and around $\sim$0.5 MByte for pp collisions at the highest assumed multiplicities~\cite{alicetpc}. 

The event rate is limited by the bandwidth of the permanent storage system. Without any further reduction or compression the ALICE TPC detector can only take central PbPb events up to 20 Hz and min.~bias pp events at a few 100 Hz. Significantly higher rates are possible by either selecting interesting (sub)events, or compressing data efficiently by modeling techniques. Both requires pattern recognition to be performed online. In order to process the detector information of 10-25 GByte/sec, a massive parallel computing system is needed, the High Level Trigger (HLT) system.

\subsection{Functionality}
\label{functionality}

The HLT system is intended to reduce the data rate produced by the detectors as far as possible to have reasonable taping costs. The key component of the system is the ability to process the raw data performing track pattern recognition in real time. Based on the extracted information, clusters and tracks, data reduction can be done in different ways:
\begin{itemize}
\item {\bf Trigger}: Generation and application of a software trigger capable of selecting interesting events from the input data stream.
\item {\bf Select}: Reduction in the size of the event data by selecting sub-events or region of interest.
\item {\bf Compression}: Reduction in the size of the event data by compression techniques. 
\end{itemize}

As such the HLT system will enable the ALICE TPC detector to run at a rate up to 200 Hz for heavy ion collisions, and up to 1 kHz for pp collisions. In order to increment the statistical significance of rare processes, dedicated triggers can select candidate events or sub-events. By analyzing tracking information from the different detectors and (pre-)triggers  online, selective or partial readout of the relevant detectors can be performed thus reducing the event rate. 
The tasks of such a trigger are selections based upon the online reconstructed track parameters of the particles, e.g. to select events which contain e$^+$e$^-$ candidates coming from quarkonium decay or to select events containing high energy jets made out collimated beams of high p$_t$ particles~\cite{hltphysics}. In the case of low multiplicity events such as for pp collisions, the online reconstruction can be used to remove pile-up events from the trigger event.

\subsection{Architecture}
\label{architecture}

The HLT system receives data from the front-end electronics. A farm of clustered SMP-nodes ($\sim$500 to 1000 nodes), based on off-the-shelf PCs and connected with a high bandwidth, low latency network provide the necessary computing power. The hierarchy of the farm has to be adapted to both the parallelism in the data flow and to the complexity of the pattern recognition. 

\begin{figure}[htb]
\begin{center}
\includegraphics[width=7cm]{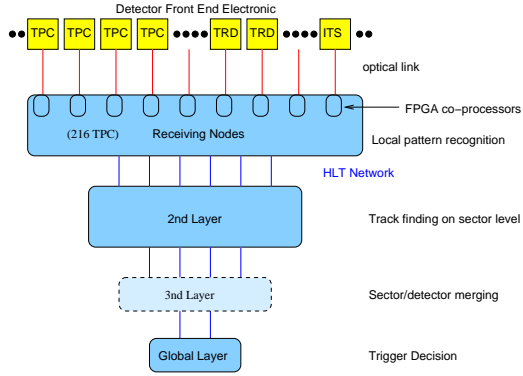}
\end{center}
\caption{Architecture of the HLT system.}
\label{hltarch}
\end{figure}

Figure~\ref{hltarch} shows a sketch of the architecture of the system. The TPC detector consists of 36 sectors, each sector being divided into 6 sub-sectors. The data from each sub-sector are transferred via an optical fiber from the detector front-end into 216 custom designed \emph{readout receiver cards} (RORCs). Each receiver node is interfaced to a RORC using its internal PCI bus. In addition to the different communication interfaces, the RORCs provide a FPGA co-processor for data intensive tasks of the pattern recognition and enough external memory to store several dozen event fractions. A hierarchical network interconnects all the receiver nodes. 

Each sector is processed in parallel, results are then merged in a higher level. The first layer of nodes receive the data from the detector and performs the pre-processing task, i.e. cluster and track seeding on the sub-sector level. The next two levels of nodes exploit the local neighborhood: track segment sending on sector level. Finally all local results are collected from the sectors or from different detectors and combined on a global level: track segment merging and final track fitting.

The farm is designed to be completely fault tolerant avoiding all single points of failure, except for the unique detector links. A generic communication framework has been developed based on the publisher-subscriber principle, which allows to construct any hierarchy of communication processing elements~\cite{pubsub}.





\section{Online Pattern Recognition}
\label{onlinepatternrecognition}

The main task of the HLT system is to reconstruct the complete event
information online. Concerning the TPC and the other tracking devices,
the particles should ideally follow helical trajectories due to the
solenoidal magnetic field of the L3 magnet, in which these detectors
are embedded. Thus we model a track by an helix with 5(+1) parameters
describing it mathematically. A track is made out of clusters. So the
pattern recognition task is extract clusters out of the raw data and
to assign them to tracks thereby determining the helix track
parameters. 

For HLT tracking, we distinguish two different approaches: the
\emph{sequential feature extraction} and the \emph{iterative feature
extraction}.  

The sequential method --corresponding to the conventional way of event
reconstruction-- first searches the cluster centroids with a
\emph{Cluster Finder} and then uses a \emph{Track Follower} on these
space points to extract the track parameters. This approach is
applicable for lower occupancy like pp and low multiplicity PbPb
collisions. However, at larger multiplicities expected for PbPb at
LHC, clusters start to overlap and deconvolution becomes necessary in
order to achieve the desired tracking efficiencies.  

For that reason, the iterative method first determines track
candidates using a suitable defined \emph{Track Candidate Finder} and
then assigns clusters to tracks using a \emph{Cluster Evaluator}
possibly deconvoluting overlapping clusters shared by different
tracks. In both cases, a helix fit on the assigned clusters finally
determines the track parameters.


In order to reduce data shipping and communicaton overhead within the
HLT, as much as possible of the \emph{local} pattern recognition will
be done on the RORC. We therefore intend to run the \emph{Cluster
Finder} or the \emph{Track Candidate Finder} directly on the FPGA
co-processor of the receiver nodes while reading out the data over the
fiber. In both cases the results, cluster centroids or track candidate
parameters, will be sent from the RORC to the host over the PCI bus. 




\section{Sequential tracking approach}
\label{seqtracker}

The classical approach of pattern recognition in the TPC is divided
into two sequential steps: Cluster finding and track finding. In the
first step the {\it Cluster Finder} reconstructs the cluster
centroids, which are interpreted as the three dimensional space points
produced by the traversing particles. The list of space points is then
passed to the {\it Track Follower}, which combines the clusters to
form track segments. A similar reconstruction chain has successfully
been used in the STAR L3 trigger~\cite{startrigger}, and thus has been
adapted  to the ALICE HLT framework.  

\subsubsection{The Cluster Finder}
\label{clusterfinder}

The input to the cluster finder is a list of above threshold timebin
sequences for each pad. The algorithm builds the clusters by matching
sequences on neighboring pads. In order to speed up the execution time
every calculation is performed {\it on-the-fly}; sequence centroid
calculation, sequence matching and deconvolution. Hence the loop over
sequences is done only once. Only two lists of sequences are stored at
every time: The current pad and the previous pad(s). For every new
sequence  the centroid position in the time direction is calculated by
the ADC weighted mean. The mean is then added to a current pad list,
and compared to the sequences in the previous. If a match is found,
the mean position in both pad and time is calculated and the cluster
list is updated. Every time a match is not found, the sequence is
regarded as a new cluster.  

In the case of overlapping clusters, a crude deconvolution scheme can
be performed~\footnote{The deconvolution can be switched on/off by a
flag of the program}. In time direction overlapping sequences are
identified by a local minimum in a sequence, and is separated by
cutting at the position of the minimum in time direction. The same
approach is being used for the  pad direction, where the cluster is
cut if there is a local minimum of the pad charge values. 

The algorithm is inherently local, as each padrow can processed
independently. This is one of the main reasons to use a circuit for
the parallel computation of the space points on the FPGA of the 
RORC~\cite{clusterfpga}.

\subsubsection{The Track Follower}
\label{trackfollower}

The tracking algorithm is based on {\it conformal mapping}. A space
point (x,y) is transformed in the following way: 
\begin{equation*}
x' = \frac{x-x_t}{r^{2}}
\end{equation*}
\begin{equation*}
y' = -\frac{y-y_t}{r^{2}}
\end{equation*}
\begin{equation}
r^2 = (x-x_t)^2 + (y-y_t)^2\,,
\end{equation}
where the reference point $(x_t,y_t)$ is a point on the trajectory of
the track. If the track is assumed to originate from the interaction
point, the reference point is replaced by the vertex coordinates. The
transformation has the property of transforming the circular
trajectories of the tracks into straight lines. Since then fitting
straight lines is easier and much faster than fitting circles (if we
neglect the changes in the weights of the points induced by conformal
mapping), the effect of the transformation is to speed up the track
fitting procedure.  

The track finding algorithm consists of a {\it follow-your-nose} where
the tracks are built by including space points close to the
fit~\cite{yepes96}. The tracks are initiated by building track
segments, and the search is starting at the outermost padrows. The
track segments are formed by linking space points which are close in
space. When a certain number of space points has been linked together,
the points are fitted to straight lines in conformal space. The tracks
are then extended by searching for clusters which are close to the
fit.  

\subsubsection{Track merging}
\label{trackmerging}

Tracking can be done either locally on every sub-sector, on the sector
level or on the complete TPC. In the first two scenarios, the tracks
have to be merged across the detector boundaries. A simple and fast
track merging procedure has been implemented for the TPC. The
algorithm basically tries to match tracks which cross the detector
boundaries and whose difference in the helix parameters are below a
certain threshold. After the tracks have been merged, a final track
fit is performed in real space. 

\begin{figure}[thb]
\centering
\includegraphics[width=7cm]{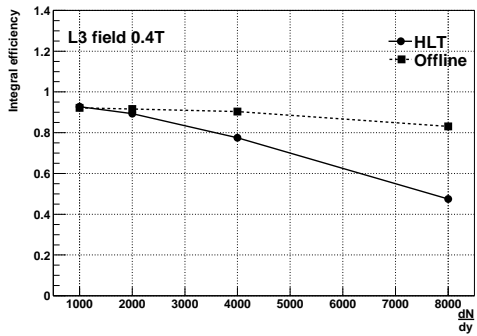}
\caption{Integral tracking efficiency for HLT online and ALIROOT offline
reconstruction as a function of different particle multiplicities 
for B=$0.4$T.}
\label{effint}
\end{figure}

\subsubsection{Tracking performance}
\label{trackperformance}

The tracking performance has been studied and compared with the
offline TPC reconstruction chain. In the evaluation the following
quantities has been defined: 

\begin{itemize}
\item {\it Generated good track} -- A track which crosses at least
40\% of all padrows. In addition, it is required that half of the
innermost 10\% of the clusters are correctly assigned.
\item {\it Found good track} -- A track for which the number of assigned
clusters is at least 40\% of the total number of padrows. In addition,
the track should not have more than 10\% wrongly assigned clusters.
\item {\it Found fake track} -- A track which has sufficient amount
of clusters assigned, but more than 10\% wrongly assigned clusters.
\end{itemize}

The tracking efficiency is the ratio of the number of {\it found good
tracks}  to the number of {\it generated good tracks}. The identical
definitions have been used both for offline and HLT for comparison.  

Figure~\ref{effint} shows the comparison of the integral efficiency of
the HLT and offline reconstruction chains for different charged
particle multiplicities for a magnetic field of B=$0.4$T. We see that
up to dN/dy of 2000 the HLT efficiency is $\ge 90$\%, but for higher
multiplicities the HLT code becomes too inefficient to be used for
physics evaluation. In this regime other approaches have to be
applied. 

\subsubsection{Timing performance}
The TPC analysis in HLT is divided into a hierarchy of processing
steps from cluster finding, track finding, track merging to track
fitting.  

\begin{figure}[hbt]
\centering
\includegraphics[height=4cm,width=7cm]{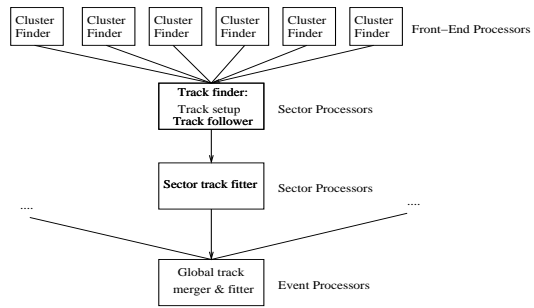}
\caption{HLT processing hierarchy for 1 TPC sector (= 6 subsectors)}
\label{hierarchy}
\end{figure}

Figure~\ref{hierarchy} shows the foreseen processing hierarchy for the
sequential approach. Cluster finding is done in parallel on each
Front-End Processor (FEP), whereas track finding and track fitting is
done sequentially on the sector level processors. The final TPC tracks
are the obtained on the event processors, where the tracks are being
merged across the sector boundaries and a final track fit is performed
(cmp. to figure~\ref{hltarch}). 

\begin{figure}[hbt]
\centering
\includegraphics[width=7cm]{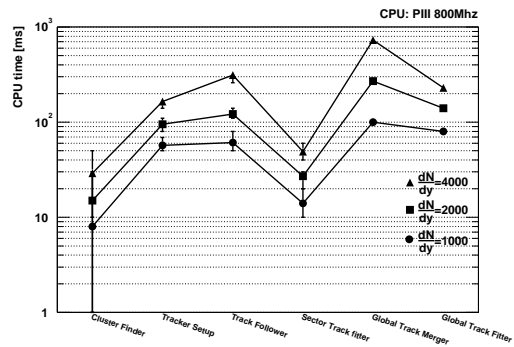}
\caption{Computing times measured on an P3 800\,MHz dual processor for 
different TPC occupancies and resolved with respect to the different 
processing steps.}
\label{timing}
\end{figure}

Figure~\ref{timing} shows the required computing time measured on a
standard reference PC~\footnote{800\,MHz Twin Pentium III, ServerWorks
Chipset, 256\,kB L3 cache} corresponding to the different processing
steps for different particle multiplicities. The error bars denote the
standard deviation of processing time for the given event
ensemble. For particle multiplicity of dN/dy=4000, about 24 seconds
are required to process a complete event, or 4800 CPUs are required to
date for the TPC alone at an event rate of 200\,Hz~\footnote{Estimate
ignores any communication and synchronization overhead in order to
operate HLT}. Table~\ref{tottime} compares the CPU time needed to
reconstruct a TPC event (dN/dy=4000) for HLT and offline.  For
offline, loading the data into memory is also included in the
measurement, while the HLT result only included the processing time as
memory  accesses are done completely transparent by the
publisher-subscriber  model~\footnote{For offline 17 \% of the time is
needed for data loading.}. 

\begin{table}
\begin{center}
\caption{Integral computing time comparison performance}
\label{tottime}
\begin{tabular}{|c|c|c|}
\hline 
dN/dy=4000 & \multicolumn{2}{|c|}{\bf CPU time (seconds)}\\
\cline{2-3} & HLT & Offline \\
\hline 
\hline
{\bf Cluster finder} & 6 & 106\\
{\bf Track finder}   & 18 & 58\\
\hline
\end {tabular}
\end{center}
\end{table}


\subsection{Iterative tracking approach}
\label{ittracking}

For large particle multiplicities clusters in the TPC start to
overlap, and deconvolution becomes necessary in order to achieve the
desired tracking efficiencies. The cluster shape is highly dependent
on the track parameters, and in particular on the track crossing
angles with the padrow and drift time. In order to properly
deconvolute the overlapping clusters, knowledge of the track
parameters that produced the clusters are necessary. For that purpose
the Hough transform is suited, as it can be applied directly on the
raw ADC data thus providing an estimate of the track parameters. Once
the track parameters are known, the clusters can be fit to the known
shape, and the cluster centroid can be correctly reconstructed. The
cluster deconvolution is geometrically local, and thus trivially
parallel, and can be performed in parallel on the rawdata.

\subsubsection{Hough Transform}

The Hough transform is a standard tool in image analysis that allows
recognition of global patterns in an image space by recognition of
local patterns (ideally a point) in a transformed parameter space. The
basic idea is to find curves that can be parametrized in a suitable
parameter space. In its original form one determines a curve in
parameter space for a signal corresponding to all possible tracks with
a given parametric form to which it could possibly belong. All such
curves belonging to the different signals are drawn in parameter
space. That space is then discretized and entries are stored in a
histogram. If the peaks in the histogram exceeds a given threshold,
the corresponding parameters are found.  

As mentioned above, in ALICE the local track model is a helix. In
order to simplify the transformation, the detector is divided into
subvolumes in pseudo-rapidity. If one restricts the analysis to tracks
originating from the vertex, the circular track in the $\eta$-volume
is characterized by two parameters: the emission angle with the beam
axis, $\psi$ and the curvature $\kappa$. The transformation is
performed from (R,$\phi$)-space to ($\psi$,$\kappa$)-space using the
following equations: 
\begin{equation*}
R = \sqrt{x^2+y^2}
\end{equation*}
\begin{equation*}
\phi = \arctan(\frac{y}{x})
\end{equation*}
\begin{equation}
\kappa = \frac{2}{R}\sin(\phi - \psi)
\end{equation}

Each ADC value above a certain threshold transforms into a sinusoidal
line extending over the whole $\psi$-range of the parameter space. All
the corresponding bins in the histogram are incremented with the
corresponding ADC-value. The superposition of these point
transformations produces a maximum at the circle parameters of the
track. The track recognition is now done by searching for local maxima
in the parameter space. 

\begin{figure}[hbt]
\centering
\includegraphics[height=7cm,width=4cm,angle=-90]{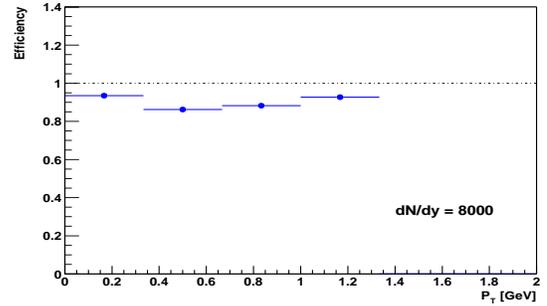}
\caption{Tracking efficiency for the Hough transform on a high occupancy event. 
The overall efficiency is above 90\%.}
\label{hougheff8000}
\end{figure}

Figure~\ref{hougheff8000} shows the tracking efficiency
for the Hough transform applied on a full multiplicity event and a
magnetic field of 0.2T. An overall efficiency above 90\% was
achieved. The tracking efficiency was taken as the
number of verified track candidates divided with the number of
generated tracks within the TPC acceptance. The list of verified track
candidates was obtained by taking the list of found local maxima and
laying a road in the rawdata corresponding to the track parameters of
the peak. If enough clusters were found along the road, the track
candidate was considered a track, if not the track candidate was
disregarded.  

However, one of the problems encountered with the Hough transform
algorithm is the number of fake tracks coming from spurious peaks in the parameter
space. Before the tracks are verified by looking into the rawdata, the
number of fake tracks is currently above 100\%. This problem has to be
solved in order for the tracks found by the Hough transform to be used
as an efficient input for the cluster fitting and deconvoluting
procedure. 

\subsubsection{Timing performance}

\begin{figure}[hbt]
\centering
\includegraphics[width=7cm]{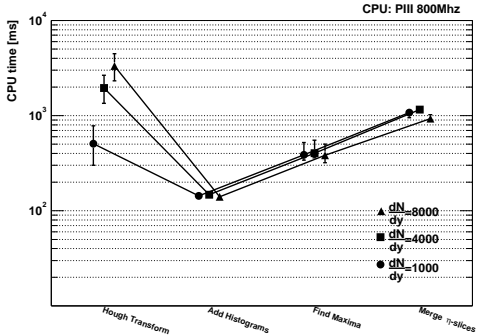}
\caption{Computation time measured on an 800\,MHz processor for different 
TPC occupancies and resolved with respect to the different processing 
steps for the Hough transform approach.}
\label{timinghough}
\end{figure}

Figure~\ref{timinghough} shows a timing measurement of the Hough based
algorithm for different particle multiplicities. The Hough transform
is done in parallel locally on each receiving node, whereas the
histogram adding, maxima finding and merging tracks across
$\eta$-slices are down sequentially on the sector level. The
histograms from the different subsectors are added in order to
increase the signal-to-noise ratio of the peaks. For particle
multiplicities of dN/dy=8000, the four steps require about 1000
seconds per events corresponding to 200,000 CPUs for 200\,Hz event
processing rate. It should be noted  that the algorithm were already
optimized but some additional optimizations are still believed to be
possible. However, present studies indicate that one should not expect
to gain more than a factor of 2 without using hardware specifics of a
given processor architecture.  

The advantage of the Hough transform is that it has a very high degree
of locality and parallelism, allowing the efficient use of FPGA
co-processors. Given the hierarchy of the TPC data analysis, it is
obvious that both the Hough transformation and the cluster
deconvolution can be performed in the receiver nodes. The Hough
transformation is particular I/O-bound as it create large histograms
that have to be searched for maxima, which scales poorly with modern
processor architectures and is ideally suited for FPGA
co-processors. Currently different ways of implementing the above
outline Hough transform in hardware are being investigated. 



\section{Data modeling and Data compression}

One of the mains goals of HLT is to compress data efficiently with a
minimal loss of physics information.

In general two modes of data compression can be considered: 
\begin{itemize}
\item {\bf Binary lossless data compression}, allowing bit-by-bit
reconstruction of the original data set.
\item {\bf Binary lossy data compression}, not allowing bit-by-bit
reconstruction of the original data, while remaining however all
relevant physical information.
\end{itemize}

Run-length encoding (RLE), Huffman and LZW are considered lossless
compression, while thresholding and hit finding operations are
considered lossy techniques that could lead to a loss of small
clusters or tail of clusters. It should be noted that data compression
techniques in this context should be considered lossless from a physics point of
view. 

Many of the state of the art compression techniques were studied on
simulated TPC data and is presented in detail in~\cite{berger02}. They
all result in compression factors of close to 2. However, the most
effective data compression can be done by cluster and track
modeling, as will be presented in the following.

\subsection{Cluster and track modeling}
From a data compression point of the view, the aim of the track
finding is not to extract physics information, but to build a data
model which will be used to collect clusters and to code cluster
information efficiently. Therefore, the pattern recognition algorithms
are optimized differently, or even different methods can be used
compared to the normal tracking.

The tracking analysis comprises of two main steps: Cluster
reconstruction and track finding. Depending on the occupancy, the
space points can be determined by a simple cluster finding or require
more complex cluster deconvolution functionality in areas of high
occupancy (see sec.~\ref{seqtracker} and \ref{ittracking}) . 
In the latter case a minimum track model may be required in
order to properly decode the digitized charge clouds into their
correct space points. 

However, in any case the analysis process is twofold, clustering and
tracking. Optionally the first step can be performed online while
leaving the tracking to offline, and thereby only recording the space
points. Given the high resolution of space points on one hand, and the
size of the chamber on the other, would result in rather large
encoding sizes for these clusters. However, taking a preliminary
zeroth order tracking into account, the space points can be encoded
with respect to their distance to such tracklets, leaving only small
numbers which can be encoded very efficiently. The quality of the
tracklet itself, with the helix parameters that would also be
recorded, is only secondary as the tracking is repeated offline with
the original cluster positions.  

\subsection{Data compression scheme}

\begin{table}
\begin{center}
\caption{Track parameters and their respective size}
\label{trackparams}
\begin{tabular}{|c|c|}
\hline
{\bf Track parameters}      & {\bf Size (Byte)} \\
\hline \hline
   Curvature               &       4 (float)\\
   X$_0$,Y$_0$,Z$_0$       &       4 (float)\\
   Dip angle,              &       4 (float)\\
   Azimuthal angle         &       4 (float)\\
   Track length            &       2 (integer)\\
   Number of clusters      &       1 (integer)\\
\hline
\end {tabular}
\end{center}
\end{table}

\begin{table}
\begin{center}
\caption{Cluster parameters and their respective size}
\label{clusterparams}
\begin{tabular}{|c|c|}
\hline
{\bf Cluster parameters}    &        {\bf Size (Bit)} \\
\hline \hline
   Cluster present         &                1\\
   Pad residual            &                9\\
   Time residual           &                9\\
   Cluster charge          &                13\\
\hline
\end {tabular}
\end{center}
\end{table}

The input to the compression algorithm is a lists of tracks and their
corresponding clusters. For every assigned cluster, the
cluster centroid deviation from the track model is calculated in both
pad and time direction. This length is quantized with respect the
given detector resolution~\footnote{The quantization steps has been
set to 0.5 mm for pad direction and 0.8 mm for time direction, which
is within the range of the intrinsic detector resolution}, and
represented by a fixed number of 
bits. In addition the total charge of the cluster is stored.
Since the cluster shape itself can be parametrized as a
function of track parameters and detector specific parameters, the
cluster widths in pad and time is not stored for every
cluster. During the decompression step, the cluster centroids are
restored, and the cluster shape is calculated based on the track
parameters. In tables~\ref{trackparams} and~\ref{clusterparams}, 
the track and cluster parameters are listed together with their respective size
being used in the compression.

\begin{figure}[thb]
\centering
\includegraphics[width=7cm]{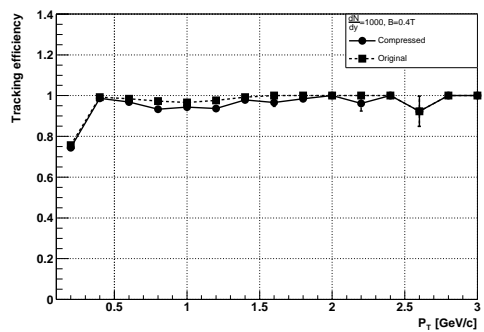}
\caption{Comparison of the tracking efficiency of the offline
reconstruction chain before and after data compression. A total loss
of efficiency of $\sim$1\% was observed.}
\label{compresseff}
\end{figure}

\begin{figure}[thb]
\centering
\includegraphics[width=7cm]{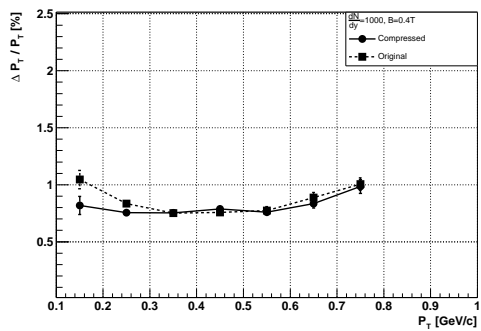}
\caption{Comparison of the p$_t$ resolution of the offline reconstruction
chain before and after data compression.}
\label{compressres}
\end{figure}

The compression scheme was applied to a simulated PbPb event with a
multiplicity of dN/dy=1000. The input tracks used were tracks
reconstructed with the sequential tracking approach (see
Fig.~\ref{effint}). The {\it remaining clusters}, or the clusters
which were not assigned to any tracks during the track finding step,
were disregarded and not stored for further analysis~\footnote{The
remaining clusters mainly originates from very low p$_t$ tracks such
as $\delta$-electrons, which could not be reconstructed by the track
finder. Their uncompressed raw data amounts to a relative size of about 20\%.}. 
A relative size of 11\% for the compressed data with respect 
to the original set was obtained. In order to evaluate the impact on the
physics observables, the data was decompressed and restored cluster processed by
the offline reconstruction chain. 

\newpage
In figure~\ref{compresseff} the
offline tracking efficiency before and after applying the compression is
compared. A total loss of $\sim$1\% in efficiency and no significant loss 
in p$_t$ resolution was observed.

However, keeping the potential gain of statistics by
the increased event rate written to tape in mind, one has to weigh the tradeoff
between the impacts on the physics observables and the costs for the
data storage.

For high occupancy events, clusters start to overlap and has to be
properly deconvoluted in order to effectively compress the data. In
this scenario, the Hough transform or another effective iterative
tracking procedure would serve as an input for the cluster
fitting/deconvolution algorithm. With a high online tracking
performance, track and cluster modeling, together with noise removal,
can reduce the data size by a factor of 10.





\section{Conclusion}

Focusing on the TPC, the sequential approach --consisting of cluster
finding followed by track finding-- is applicable for pp and low
multiplicity PbPb data up to dN/dy of 2000 to 3000 with more than
90\% efficiency. The timing results indicate that the desired
frequency of 1KHz for pp and 200 Hz for PbPb can be achieved. For
higher multiplicities of dN/dy $\ge$ 4000 the iterative approach using
the Circle Hough transform for primary track candidate finding shows
promising efficiencies of around 90\% but with high computational
costs. 

By compressing the data using data modeling, results show that one can
compress data of up to 10\% relative to the original data with a
very small impact on the tracking efficiency and the P$_t$
resolution. 



\end{document}